\documentclass[]{aa} 
\usepackage{graphicx}
\newcommand{\iz}{I\,Zw\,18}
\newcommand{\kms}{km\,s$^{-1}$}

\begin{document}
%
%
\title{{\sl FUSE} observations of the H\,{\sc I} interstellar gas of I\,Zw\,18}
\titlerunning{\sl {FUSE} observations of I\,Zw\,18}
%

   \author{
A.~Lecavelier des Etangs \inst{1}
          \and
J.-M.~D\'esert \inst{1}
          \and
D.~Kunth \inst{1}
          \and
A.~Vidal-Madjar \inst{1}
          \and
G.~Callejo \inst{1,2}
          \and
R.~Ferlet \inst{1}
          \and
G.~H\'ebrard \inst{1}
          \and
V.~Lebouteiller \inst{1}
 }

   \offprints{A. Lecavelier des Etangs,
   \email{lecaveli@iap.fr}}

   \institute{Institut d'Astrophysique de Paris, CNRS, 98 bis Bld Arago,
              F-75014 Paris, France
         \and
             LERMA, Observatoire de Paris - Section de Meudon,
                                 5 Place Jules Janssen
                            92195 Meudon cedex, France
  }

   \date{Received 28 May 2003 / Accepted 20 September 2003}

   \abstract{
We present the analysis of {\sl FUSE} observations of the
metal-deficient dwarf galaxy \iz . We measured column densities of
H\,{\sc i}, N\,{\sc i}, O\,{\sc i}, Ar\,{\sc i}, Si\,{\sc ii}, and
Fe\,{\sc ii}. The O\,{\sc i}/H\,{\sc i} ratio ($\log$(O\,{\sc
i}/H\,{\sc i})=$-4.7^{+0.8}_{-0.6}$) is consistent with the O/H
ratio observed in the H\,{\sc ii} regions (all uncertainties are
2-$\sigma$). If the oxygen is depleted in the H\,{\sc i} region
compared to the H\,{\sc ii} regions, the depletion is at most
0.5~dex. This is also consistent with the $\log$(O/H) ratios $\sim -5$
measured with {\sl FUSE} in the H\,{\sc i} regions of other blue
compact dwarf galaxies. With $\log$(N\,{\sc i}/O\,{\sc
i})=$-2.4^{+0.6}_{-0.8}$, the measured N\,{\sc i}/O\,{\sc i} ratio
is lower than expected for primary nitrogen. The determination of
the N\,{\sc ii} column density is needed to discriminate between a
large ionization of N\,{\sc i} or a possible nitrogen deficiency.
The neutral argon is also apparently underabundant, indicating
that ionization into Ar\,{\sc ii} is likely important. The
column densities of the other $\alpha$-chain elements Si\,{\sc ii}
and Ar\,{\sc i} favor the lower edge of the permitted range of
O\,{\sc i} column density, $\log (N_{\rm cm^{-2}}$(O\,{\sc
i}))$\sim 16.3$.
   }

   \maketitle
%

\section{Introduction}
\label{introduction}

\iz\ (Mrk 116) is a blue compact dwarf galaxy with the smallest known abundance
of heavy elements as measured from nebular emission lines.
The observed emission lines originate from a pair of bright
H\,{\sc ii} regions in which a strong
burst of star formation is taking place. In these H\,{\sc ii} regions,
the oxygen abundance is measured
to be only $\sim$ 1/50 the solar value (Izotov et al.~\cite{Izotov1999}).
The interstellar gas ionized by recently formed massive stars is surrounded by
an envelope of neutral hydrogen (H\,{\sc i}). This envelope has been
extensively observed through 21cm radio maps
which give a total mass of $M_{\rm HI} \sim 5\times 10^{7}$M$_\odot$
(Lequeux \& Viallefond~\cite{Lequeux1980};
Viallefond et al.~\cite{Viallefond1987}; van Zee et al.~\cite{vanZee1998}).

The composition of this H\,{\sc i} gas has been debated.
For instance, it had been suggested that it may
be a significant reservoir of molecular hydrogen
representing a significant fraction of the dark matter
(Lequeux \& Viallefond~\cite{Lequeux1980}). In fact, early {\sl FUSE}
(Far Ultraviolet Spectroscopic Explorer; Moos et al.~\cite{Moos2000})
observations proved that diffuse molecular hydrogen is very scarce with
a column density ratio $N_{\rm H_2}/N_{\rm HI}< 10^{-6}$ (Vidal-Madjar et al.~\cite{Vidal-Madjar2000}).

Also there is a disagreement about the metallicity of the
neutral gas. Kunth \& Sargent (\cite{Kunth1986}) postulated the importance
of the ``self-pollution'' of heavy elements in the H\,{\sc ii} regions
by the present star burst. They suggested that the
neutral gas might be more primordial than the H\,{\sc ii} regions
and close to zero metallicity, providing thus a new interesting site to determine
the primordial abundances of key elements like deuterium and helium
(Kunth et al.~\cite{Kunth1995}).
HST has been used to test this hypothesis
through the observation of the O\,{\sc i} line at 1302~\AA.
O\,{\sc i} is indeed one of the major tracer of the metallicity.
The ionization potential
of O\,{\sc i} is very close to that of H\,{\sc i} and
the charge exchange between O\,{\sc ii} and H\,{\sc i}
is very efficient. Hence the coupling of the oxygen
and hydrogen ionization fractions is very strong.
In the neutral gas, the O\,{\sc i}/H\,{\sc i} ratio can therefore
be considered as a very good proxy for the O/H ratio and hence for
the metallicity.

Unfortunately the O\,{\sc i} line at 1302~\AA\ has a very strong
oscillator strength ($f_{1302}=51.9\times 10^{-3}$). With an
intrinsic line width of $b\sim 18$~\kms\ in the case of \iz , and
for the metallicity of the ionized gas $Z_{\rm HII\ IZw18}\sim
1/50~Z_{\odot}$, this O\,{\sc i} line is strongly saturated
already at $N_{\rm HI}\ga 2\times 10^{20}$cm$^{-2}$, while the
observed $N_{\rm HI}$ is $\ge 2 \times 10^{21}$cm$^{-2}$. As a
result, despite deep HST observations, the oxygen abundance is
still an open question (Kunth et al.~\cite{Kunth1994}; Pettini et
al.~\cite{Pettini1995}). Using also the O\,{\sc i} line at
1302~\AA, Thuan et al. (\cite{Thuan1997}) claimed that the H\,{\sc
i} envelope of the dwarf galaxy SBS~0335-052 was extremely metal
deficient with an O\,{\sc i}/H\,{\sc i} ratio $\la 3 \times
10^{-7}$. However in that same galaxy the O\,{\sc i}/H\,{\sc i}
ratio was subsequently found to be much higher using {\sl FUSE}
observations of O\,{\sc i} lines at shorter wavelengths and with
smaller oscillator strengths (Lecavelier des Etangs et
al.~\cite{Lecavelier2002}; Thuan et al.~\cite{Thuan2003}). In
fact, the hypothesis of an O\,{\sc i} line at 1302~\AA\
unsaturated was simply erroneous.

It is clear that the far-UV is unique for offering
a large collection of lines of different elements
with a wide range of oscillator strengths.
We therefore observed \iz\ with {\it FUSE} to determine the
abundances of various species, and constrain the
metallicity and the history of elements in one of the most extreme
blue compact dwarf galaxy.

In Sect.~2 we describe the observations.
In Sect.~3 we present the analysis of the absorption lines against
the stellar continuum from the blue massive stellar clusters. Results
are presented in Sect.~4 and discussed in Sect.~5.
A comparison with {\sl FUSE} observations of
other metal-deficient dwarf galaxies
will be part of a forthcoming paper
(Lebouteiller et al.~\cite{Lebouteiller2003}).

\section{Observation}

\iz\ was observed with {\sl FUSE} (900-1200~\AA , see Moos et
al.~\cite{Moos2000}) and Sahnow et al.~\cite{Sahnow2000}) through
the LWRS aperture ($30\arcsec$$\times$$30\arcsec$) for a total of
31\,650 seconds on November 27, 1999 (Program P1980102) and
63\,450 seconds on February 11, 2002 (Program P1080901). During
the early observations of November, 1999 (Vidal-Madjar et
al.~\cite{Vidal-Madjar2000}), the {\sl SiC} channels acquisition
failed and the resulting spectrum was limited to wavelengths
longward of about 1000~\AA. Therefore, the {\sl SiC} wavelength
range ($\sim$900--1000~\AA) was only exposed for 63\,450 seconds.
The data have been reprocessed with the version 2.0.1 of the {\tt
CALFUSE} pipeline. The output of the pipeline is a total of
38~sub-exposures which have been aligned and co-added, resulting
in a set of four independent spectra, one for each {\sl FUSE}
channel (two {\sl LiF} spectra and two {\sl SiC} spectra).

   \begin{figure*}[tbp]
   \centering
   \resizebox{\textwidth}{!}
             {\includegraphics[]{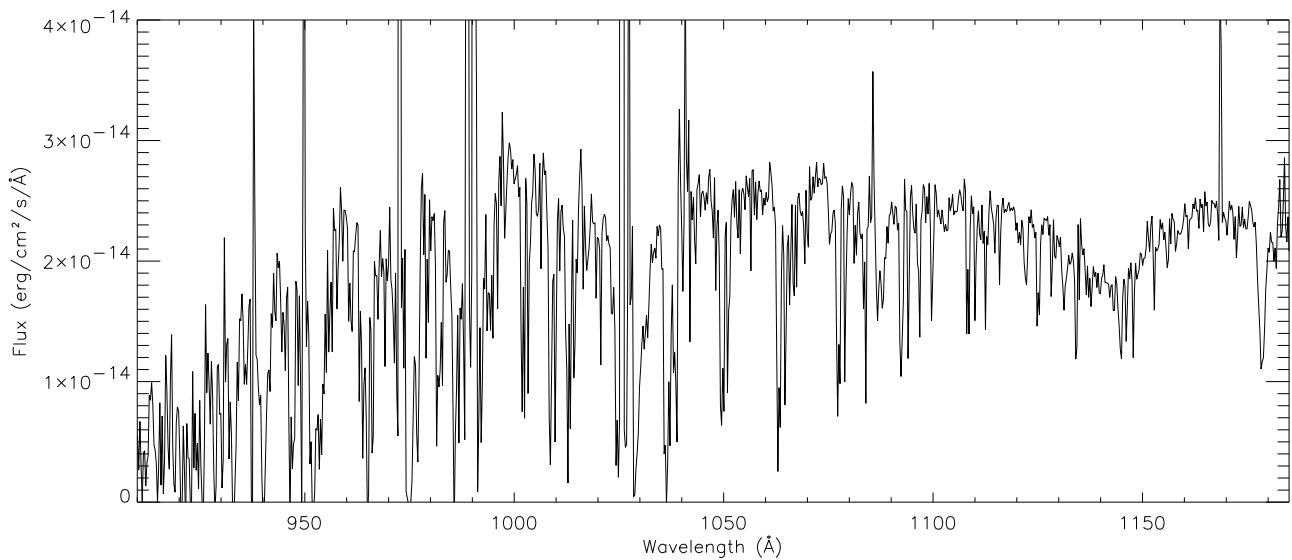}}
      \caption{Plot of the full {\sl FUSE} spectrum of \iz . It is obtained
through the addition of all the channels,
rebinned by 0.3~\AA. In addition to the Earth airglow emission lines
and an artifact dip around 1145~\AA, many absorption lines from
the Milky Way and \iz\ are clearly visible. For instance, even at this low
resolution, we can see the H\,{\sc i} Lyman $\beta$, $\gamma$, $\delta$,
and $\epsilon$ lines from \iz\ at 1030, 975, 952, and 940~\AA, respectively.}
    \label{spectrum}
    \end{figure*}

Many absorption lines are clearly detected (Fig.~\ref{spectrum}).
For a given element, they consist of three main absorbing
components at heliocentric radial velocities: $-160$~\kms, $\sim
0$~\kms, and $750$~\kms\ (Vidal-Madjar et
al.~\cite{Vidal-Madjar2000}). These can easily be identified
respectively with a known Galactic high velocity cloud at
$-160$~\kms, the interstellar medium of the Milky Way at low
radial velocity, and \iz\ itself at $750$~\kms\ corresponding to a
Doppler shift of about $\Delta \lambda\sim 2.5$~\AA\ at 1000~\AA .
The high velocity cloud is detected in H\,{\sc i}, C\,{\sc ii},
N\,{\sc i}, O\,{\sc i}, Fe\,{\sc ii}, and Fe\,{\sc iii} lines. The
Galactic clouds appear in C\,{\sc i}, C\,{\sc i}*, C\,{\sc ii},
C\,{\sc iii}, N\,{\sc i}, O\,{\sc i}, Si\,{\sc ii}, Ar\,{\sc i},
Fe\,{\sc ii}, Fe\,{\sc iii}, and also in molecular hydrogen
(H$_2$). H$_2$ lines are detected up to the $J=5$ level. These
electronic transitions of H$_2$ only show up in the Galactic
medium. Finally \iz\ absorption is detected in H\,{\sc i}, C\,{\sc
ii}, C\,{\sc iii}, N\,{\sc i}, O\,{\sc i}, S\,{\sc iii}, Ar\,{\sc
i}, Si\,{\sc ii}, Fe\,{\sc ii}, and Fe\,{\sc iii}.

In addition to these three main components, O\,{\sc vi} lines are detected
around 1032~\AA\ and 1037~\AA .
They correspond to coronal gas with
radial velocities between $-125$~\kms\ and $+110$~\kms , originating in the
Galactic halo (Savage et al.~\cite{Savage2003}; Sembach et al.~\cite{Sembach2003};
Wakker et al.~\cite{Wakker2003}).
As already shown by Vidal-Madjar et al. (\cite{Vidal-Madjar2000}), no line from H$_2$
is observed at the radial velocity of \iz . This absence of diffuse
molecular hydrogen is also observed in the blue compact dwarf Mrk~59
(Thuan et al.~\cite{Thuan2002}) and is well explained by the low abundance of dust grains,
the high ultraviolet flux, and the low density of the H\,{\sc i} gas.

\section{Method}

Column densities have been calculated through profile fitting
using the {\tt Owens} procedure developed by Martin Lemoine and
the {\sl FUSE} French team. This code returns the most likely
values of many free parameters like the Doppler widths and column
densities through a $\chi^2$ minimization of the difference
between the observed and computed profiles. The latest version of
this code is particularly suited to the characteristics of {\sl
FUSE} spectra. For example,  it allows for a variation of the
background level, for an adjustable line spread function as a
function of the wavelength domain, and for shifts in wavelength
scale. These are taken as free parameters which depend on the
wavelength region and are determined by the $\chi^2$ minimization.
Special attention was paid to the background residual which is not
negligible for this faint target. It is found to lie between $-6$
and $+4\times 10^{-15}$erg~cm$^{-2}$~s$^{-1}$~\AA$^{-1}$. The line
spread function is mainly constrained by the unresolved Galactic
H$_2$ absorption lines. The resolving power is found to be about
$R=\Delta \lambda/\lambda \approx 10000$, corresponding to the
size of the extended object within the spectroscope slit
(Vidal-Madjar et al.~\cite{Vidal-Madjar2000}).

Values of column densities are given in Table~\ref{N} (althrough the paper,
error bars are 2-$\sigma$).
The error bars are estimated by the classical method of the $\Delta \chi^2$
increase of the $\chi^2$ of the fit;
see H\'ebrard et al.~(\cite{Hebrard2002}) for a full discussion of the fitting
method and error estimation with the {\tt Owens} code.
These error bars include
the uncertainties in the continuum, intrinsic line widths ($b$),
and the instrumental line spread function.
It has to be noted that the use of different lines with different
oscillator strengths of the same species allows us to constrain
all these quantities. In particular, fits of saturated lines
constrain the line widths; the instrumental line spread function
is mainly constrained through the saturated and very narrow lines
from H$_2$ in the Milky Way.
The final results are obtained from a simultaneous self-consistent
fit to all the data.
We obtain $b_{\rm IZw18}=17.3\pm 5.0 $~km~s$^{-1}$ (2-$\sigma$)
for the width of the lines at the \iz\ systemic velocity $v_{\rm IZw18}=750$~km~s$^{-1}$.
This is consistent with 21~cm observations (van Zee et al. \cite{vanZee1998}), which
also show that most of the observed lines width can be attributed
to the velocity gradient of the neutral gas in front of the UV-bright stars.

   \begin{figure}[tbp]
   \centering
   \includegraphics[width=\columnwidth]{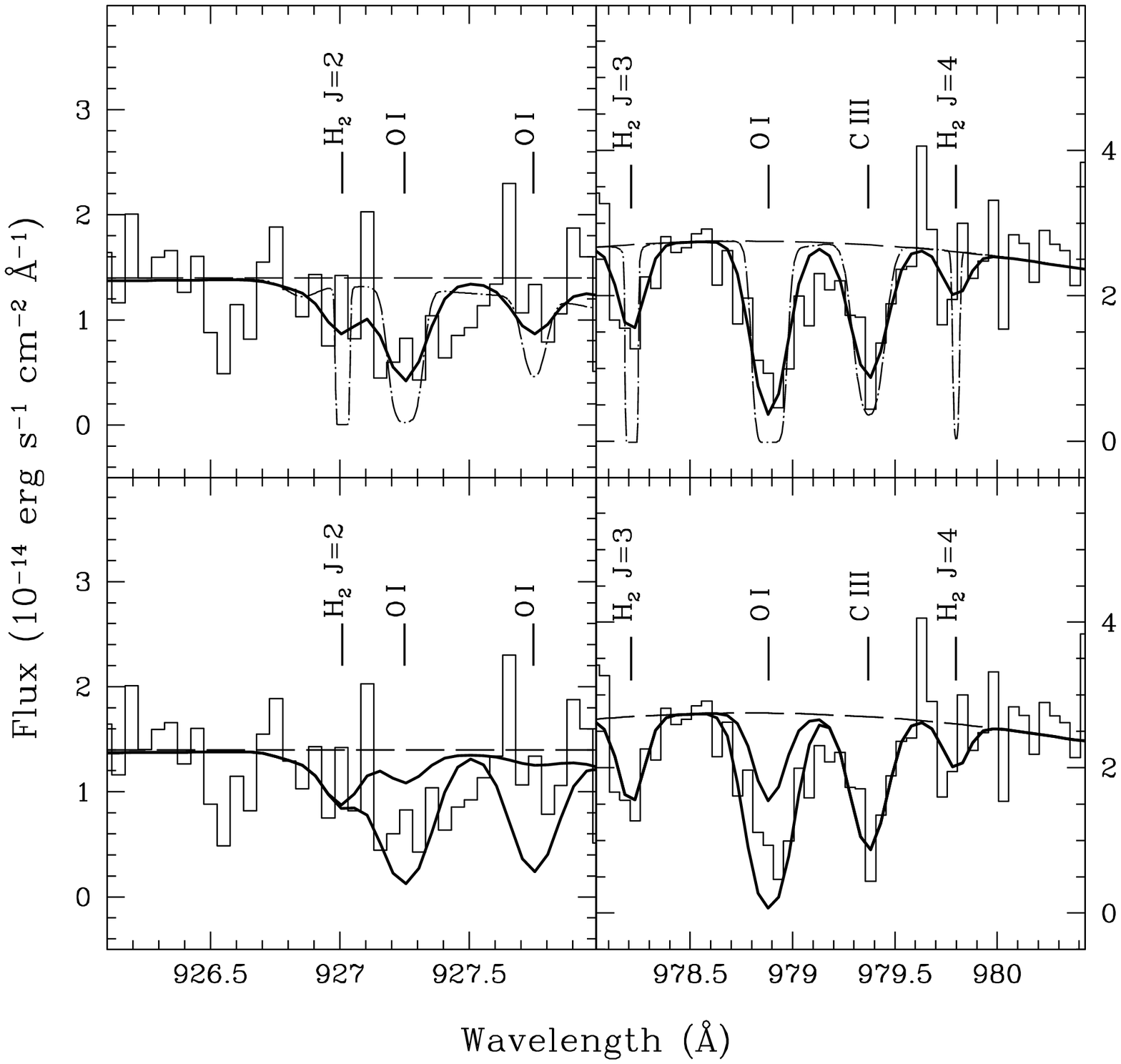}
      \caption{
Sample of O\,{\sc i} lines detected at the redshift of \iz\ (750 km\,s$^{-1}$).
The H$_2$ lines are from the Milky Way. The C\,{\sc iii} line is
at a redshift of 720 km\,s$^{-1}$.
The histogram shows the data. The upper panels show
the final fit (thick line) together with
the theoretical spectrum before convolution with the instrumental
line spread function (thin dot-dashed line).
The O\,{\sc i} line at 927.7~\AA\ is not saturated,
the line at 927.2~\AA\ is barely saturated, while the one at 978.9~\AA\ is saturated.
The bottom panels show the same data with the theoretical profiles
using an O\,{\sc i} column density
reduced and increased by a factor of 10 (i.e. $\log N$(OI)=15.6 and 17.6,
respectively). Such column densities are clearly excluded. With $\log N$(OI)=15.6,
the line at 927.2~\AA\ would not be detectable, and the line at 978.9~\AA\ would
be much fainter than observed.  With $\log N$(OI)=17.6, the line at 927.7~\AA\ would
be much deeper than observed.
}
    \label{fit_OI}
    \end{figure}

   \begin{figure}[htbp]
   \centering
   \includegraphics[width=\columnwidth]{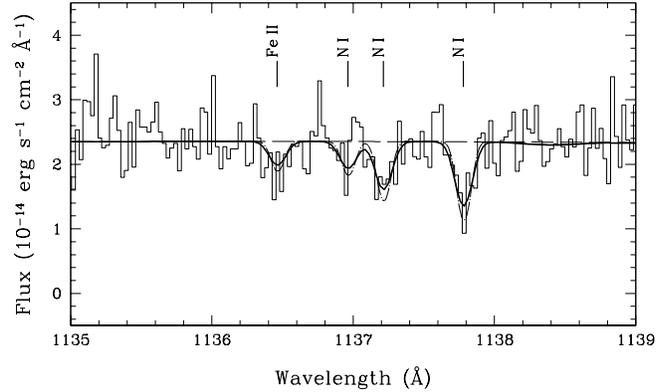}
      \caption{Plot of the N\,{\sc i} 1134~\AA\ triplet in \iz .
The thick line is the fit to the data (histogram). The equivalent widths of the
observed lines are proportional to their oscillator strengths, indicating these
lines are not saturated. The mean column density of
the gas is thus well constrained,
even in the case of multiple lines of sight toward many bright stars
(see Sect.~\ref{Multiple lines of sight}).
}
    \label{fit_NI}
    \end{figure}

\section{Results}

\subsection{Column densities}

   \begin{table}[htbp]
      \caption[]{Column densities measured in the {\sl FUSE} spectrum of \iz .}
         \label{N}
         \begin{tabular}{lll}

            \hline
            \hline
            \noalign{\smallskip}
Species & $\log_{10}$ ($N_{\rm cm^{-2}}$) & 2-$\sigma$ Error bars \\
            \noalign{\smallskip}
            \hline
            \noalign{\smallskip}
H\,{\sc i}    & 21.30 & $^{+0.10}_{-0.10}$   \\
            \noalign{\smallskip}
N\,{\sc i}    & 14.22 & $^{+0.14}_{-0.17}$   \\
            \noalign{\smallskip}
O\,{\sc i}    & 16.60 & $^{+0.80}_{-0.55}$   \\
            \noalign{\smallskip}
Ar\,{\sc i}    & 13.45 & $^{+0.20}_{-0.25}$   \\
            \noalign{\smallskip}
Si\,{\sc ii}    & 14.80 & $^{+0.25}_{-0.25}$   \\
            \noalign{\smallskip}
Fe\,{\sc ii}    & 14.60 & $^{+0.12}_{-0.10}$   \\
            \noalign{\smallskip}
            \hline
         \end{tabular}
   \end{table}

   \begin{table}[htbp]
      \caption[]{Oscillator strengths of a sample of useful N\,{\sc i} and O\,{\sc i} lines}
         \label{f}
         \begin{tabular}{lcrcl}

            \hline
            \hline
            \noalign{\smallskip}
& $\lambda_0$   & $\lambda_{\rm obs}$ &  $f$  & comment\\
 & (\AA )& (\AA )\\
            \noalign{\smallskip}
            \hline
            \noalign{\smallskip}
N\,{\sc i} & 953.42 & 955.8 & $1.32\times 10^{-2}$ & not detected\\
N\,{\sc i} & 953.65 & 956.0 & $2.50\times 10^{-2}$ & ''\\
N\,{\sc i} & 953.92 & 956.3 & $3.48\times 10^{-2}$ & faint detection\\
N\,{\sc i} & 954.10 & 956.5 & $0.68\times 10^{-2}$ & not detected\\
\\
N\,{\sc i} & 963.99 & 966.4 & $1.48\times 10^{-2}$ & not detected\\
N\,{\sc i} & 964.63 & 967.0 &$0.94\times 10^{-2}$ & ''\\
N\,{\sc i} & 965.04 & 967.4 &$0.40\times 10^{-2}$ & ''\\
\\
N\,{\sc i} & 1134.17 & 1137.0 & $1.52\times 10^{-2}$ & not detected \\
N\,{\sc i} & 1134.41 & 1137.2 & $2.97\times 10^{-2}$ & detected, \\
 & & &                                               & not saturated\\
N\,{\sc i} & 1134.98 & 1137.8 & $4.35\times 10^{-2}$ & ''\\
            \noalign{\smallskip}
            \hline
            \noalign{\smallskip}
O\,{\sc i} &  924.95 & 927.2 & $1.59\times 10^{-3}$ & detected, \\
 & & &                                               & slightly saturated\\
O\,{\sc i} &  925.44 & 927.7 & $0.35\times 10^{-3}$ & not detected\\
O\,{\sc i} &  976.45 & 978.9 & $3.31\times 10^{-3}$ & detected, saturated\\
O\,{\sc i} & 1039.23 & 1041.8 & $9.20\times 10^{-3}$ & strongly saturated \\
 & & &                                               & + airglow\\
O\,{\sc i} & 1302.17 & 1305.4 & $51.9\times 10^{-3}$ & strongly saturated\\
 & & &                                               &  ({\sl HST} observations)\\
            \noalign{\smallskip}
            \hline
         \end{tabular}
   \end{table}

\subsubsection{H\,{\sc i}}

The H\,{\sc i} column density in \iz\ is derived from all the
available lines of the Lyman series. We avoided the blue wing of
the Lyman~$\beta$ line and the Lyman~$\gamma$ line, both
contaminated by the terrestrial airglow. We fitted the H\,{\sc i}
Lyman series including the blue wing of the Lyman $\beta$ line,
the Lyman $\delta$, $\epsilon$, $\zeta$, and $\eta$ lines. Note
that the resulting H\,{\sc i} column density is mainly constrained
by the Lyman $\beta$ line. We find $N$(H\,{\sc i})$\approx (2.0\pm
0.5)\times 10^{21}$cm$^{-2}$ (2-$\sigma$ uncertainties).
This value is consistent with $N$(H\,{\sc i})$\approx 2.1\times
10^{21}$cm$^{-2}$ found by Vidal-Madjar et al.
(\cite{Vidal-Madjar2000}) from the early {\sl FUSE} observations.
It is also comparable to $N$(H\,{\sc i})$\approx 3.5\times
10^{21}$cm$^{-2}$ obtained from {\sl HST/GHRS} observations (Kunth
et al.~\cite{Kunth1994}) and to the peak column density of
$3.0\times 10^{21}$cm$^{-2}$ of the H\,{\sc i} 21~cm observations
(van Zee et al.~\cite{vanZee1998}). However with the very
different size of the {\sl GHRS} aperture and the VLA synthesized
beam, these last values could have been more different, as it is
the case for the higher column density of about $2\times
10^{22}$cm$^{-2}$ found with the smaller aperture of {\sl
HST/STIS} (Brown et al.~\cite{Brown2002}).

By comparison with H\,{\sc i} 21~cm maps, the measured H\,{\sc i}
column density gives a clue on the location of the absorber.
H\,{\sc i} column densities of the order of $10^{21}$cm$^{-2}$
correspond to the central part of \iz. Such high column densities
can be found only within the optical extend of the galaxy. The
absorber is thus likely located in the very central part of the
galaxy, close to the UV-bright stars and their associated H\,{\sc
ii} regions.

\subsubsection{C\,{\sc ii} and N\,{\sc ii}}

The only strong C\,{\sc ii} line in the {\sl FUSE} wavelength range is
the line at $\lambda_0$=1036.3~\AA . Unfortunately, at the
\iz\ redshift this line is partially blended with an O\,{\sc i} airglow emission.
It is therefore difficult to obtain a meaningful estimate of the
C\,{\sc ii} column density.

At the redshift of \iz , the N\,{\sc ii} lines from the ground and
excited levels around 1084~\AA\ fall in the {\sl FUSE} detector
gap. These lines are only detected at the upper edges of the {\sl
SiC} detectors, which are inefficient at these wavelengths.
N\,{\sc ii} column densities cannot be estimated using the present
{\sl FUSE} observations.

\subsubsection{C\,{\sc iii}, S\,{\sc iii}, and Fe\,{\sc iii}}

Absorption lines from highly ionized species are also detected at the
\iz\ redshift : C\,{\sc iii}
($\lambda_0$=977.0~\AA ), S\,{\sc iii} ($\lambda_0$=1012.5~\AA),
and Fe\,{\sc iii} ($\lambda_0$=1122.5~\AA).
The S\,{\sc iii} line is blended with Galactic H$_2$, while the Fe\,{\sc iii}
line is bracketed between Fe\,{\sc ii} lines from \iz\ and from the Milky Way.
The C\,{\sc iii} line is clean and at a
slightly different redshift of about 720~\kms\ at 979.36~\AA\
(Fig.~\ref{fit_OI}).
Voigt profiles give the following
column densities: $N$(C\,{\sc iii})$\approx 4\times 10^{13}$cm$^{-2}$,
$N$(S\,{\sc iii})$\approx 1\times 10^{14}$cm$^{-2}$, and
$N$(Fe\,{\sc iii})$\approx 3\times 10^{14}$cm$^{-2}$.

However, we cannot exclude a blend of those lines with stellar
lines. In effect, we find a large intrinsic width for the Fe\,{\sc
iii} line: $b_{\rm FeIII} \ga 90$~\kms . Consequently, the derived
column densities must be considered as upper limits to the true
values.

\subsubsection{O\,{\sc i}}

O\,{\sc i} is a tracer of the metallicity. The present new {\sl
FUSE} data provides the unprecedented opportunity to shed light on
a possible metallicity difference between the ionized and the
neutral gas. Therefore the fitting of the O\,{\sc i} lines has
been performed with great care. Many of the O\,{\sc i} lines are
blended either with Galactic H\,{\sc i} lines (O\,{\sc i} lines at
rest-wavelengths $\lambda_0$=916.8~\AA , 918.0~\AA , 930.2~\AA ,
and 937.8~\AA) or with Galactic H$_2$ lines (lines at
$\lambda_0$=921.9~\AA , 929.5~\AA , 936.6~\AA , 971.7~\AA, and
988.7~\AA ) or with Galactic N\,{\sc i} ($\lambda_0$=950.9~\AA),
or with the \iz\ H\,{\sc i} Lyman~$\delta$ line
($\lambda_0$=948.7~\AA). The O\,{\sc i} line at
$\lambda_0$=1039~\AA\ is strongly saturated and contaminated by a
terrestrial airglow line, it is thus of no use for the column
density determination. Finally we end up with the detection of
three useful O\,{\sc i} lines ($\lambda_0$=924.95~\AA ,
925.44~\AA\ and 976.45~\AA , see Table~2). We find $\log (N_{\rm
cm^{-2}}$(O\,{\sc i}))=16.6$^{+0.8}_{-0.55}$ (2-$\sigma$
uncertainties, Fig.~\ref{fit_OI}).

As a check, we also derived the O\,{\sc i} column density by using
the line at 1302~\AA\ observed with the {\sl GHRS} spectrograph
and the {\sl LSA} aperture of the {\sl HST} on April 22, 1992.
Individual {\sl GHRS} `FP-split' exposures have been aligned by
cross-correlation and co-added. The profile fitting has been done
simultaneously for the {\sl HST} and {\sl FUSE} O\,{\sc i} lines
(see Table~2). Here, only the O\,{\sc i} lines have been used to
constrain the absorbing cloud parameters like the intrinsic line
width $b$ which is taken as a free parameter. We found $\log
(N_{\rm cm^{-2}}$(O\,{\sc i}))=16.7$^{+0.6}_{-0.7}$, which is
consistent with the value derived from {\sl FUSE} data alone
(Table~1). However, to be conservative, as the
$\lambda_0$=1302~\AA\ line is strongly saturated and the {\sl
GHRS} aperture is very different from the {\sl FUSE} aperture, we
will adopt the value obtained with {\sl FUSE} alone for further
discussion.

\subsubsection{N\,{\sc i}}

Lines of the N\,{\sc i} triplet at 1134~\AA\ are easily detected
(Fig.~\ref{fit_NI}). It is clear from the plot that the equivalent
widths of the lines nicely follow the oscillator strengths
(Table~2), indicating that these lines are not saturated. The
simultaneous fit of these lines and other multiplets (at
$\lambda_0$$\approx$953~\AA\ and $\lambda_0$$\approx$964~\AA )
gives an accurate estimate of the N\,{\sc i} column density :
$N($N\,{\sc i}$)= (1.7\pm 0.6) \times 10^{14}$~cm$^{-2}$ 
(2-$\sigma$).

\subsubsection{Ar\,{\sc i}, Si\,{\sc ii}, and Fe\,{\sc ii}}

Two Ar\,{\sc i} lines are present in the {\sl FUSE} wavelength
range. The one at $\lambda_0$=1066~\AA\ is barely detected, while
the strongest one at $\lambda_0$=1048~\AA\ (4-time larger
oscillator strength), is clearly detected but blended with a
Galactic H$_2$ ($J=1$) line (see Fig.~1 in Vidal-Madjar et
al.~\cite{Vidal-Madjar2000}). Fortunately this Galactic H$_2$
($J=1$) is also detected in many other Lyman and Werner bands.
Thanks to the {\tt Owens} code, from a simultaneous fit to the
complete set of data including all the H$_2$ and Ar\,{\sc i}
lines, we can estimate the argon column density to be $N($Ar\,{\sc
i}$)= 2.8^{+1.7}_{-1.2} \times 10^{13}$~cm$^{-2}$. These error
bars include the uncertainties in the H$_2$ ($J=1$) column density
and in the instrumental profile. Consequently, they include also
the uncertainties in the perturbation introduced by the blend with
the H$_2$ line.

The Si\,{\sc ii} column density is derived from the strong line at
$\lambda_0$=989.9~\AA , and the line at $\lambda_0$=1020.7~\AA\
which is also blended with a Galactic H$_2$ line (see Fig.~1 in
Vidal-Madjar et al.~\cite{Vidal-Madjar2000}). We find $N($Si\,{\sc
ii}$)= 6.3^{+4.9}_{-2.8} \times 10^{14}$~cm$^{-2}$.

The Fe\,{\sc ii} column density towards the UV bright stars of
\iz\ is well constrained by a large set of Fe\,{\sc ii} lines that
span a large range of oscillator strengths. From the simultaneous
fit of all these lines we find $N($Fe\,{\sc ii}$)=
4.0^{+1.3}_{-0.8} \times 10^{14}$~cm$^{-2}$.

\subsection{Multiple lines of sight}
\label{Multiple lines of sight}

The analysis of absorption lines from galaxies raises a specific
problem. The observed spectrum is the sum of the fluxes coming
from thousands of hot stars through different absorbers with
varying column densities and radial velocities.

For instance the O\,{\sc i} line at $\lambda_0=1039.23$~\AA\
observed toward Mrk~59 is clearly larger than the line spread
function and has a squared shape with a maximum absorption of
about 25\% (see Fig.~6 in Thuan et al.~\cite{Thuan2002}). To fit
that line, Thuan et al. (\cite{Thuan2002}) added multiple profiles
with radial velocities distributed over 100~\kms\ each having
$b$=7~\kms . However, such a particular velocity structure is not
needed for \iz . In effect, a careful inspection of the saturated
O\,{\sc i} line at $\lambda_{\rm obs}$=978.9~\AA\ shows that it
goes almost to the zero level and that its width is not much
larger than the width of the line before convolution with the
instrumental profile (Fig.~\ref{fit_OI}). All that indicates that
the dispersion of the radial velocities of the individual clouds
in the multiple lines of sight is not significantly larger than
the intrinsic line width. The H\,{\sc i} Lyman lines also shows
that a special profile fitting procedure taking the multiple lines
of sight into account is not needed for \iz . Single Voigt
profiles are sufficient to explain the lines structures. Thus the
derived column densities for \iz\ are barely affected by the
assumption of a single absorbing cloud in a single line of sight.

In addition, in the case of different lines of sight with similar
column densities and intrinsic line widths but different radial
velocities, the absorption profiles are well reproduced with
single profiles convolved with a larger line spread function. All
the lines observed towards \iz, coming from \iz , the Milky Way or
the high velocity cloud at -160~\kms, can be fitted with the same
instrumental line spread function. This reinforces the belief that
the dispersion of the physical properties of the individual clouds
in the different lines of sight toward \iz\ is limited.

Finally, for unsaturated lines like those of N\,{\sc i}
(Fig.~\ref{fit_NI}), column densities obtained from a single Voigt
profile are not affected by the complexity of the line of sight.
For saturated lines like those of O\,{\sc i}, a single Voigt
profile can underestimate the saturation and thus column densities
(see Thuan et al.~\cite{Thuan2002}). However, in the case of
O\,{\sc i}, a larger column density would result in an O/H ratio
larger in the H\,{\sc i} gas than in the H\,{\sc ii} regions; this
seems unlikely (see Sect~\ref{Discussion}).

\section{Discussion}
\label{Discussion}

\subsection{The O/H ratio and the metallicity of the H\,{\sc i} gas in \iz }

A main issue which can be addressed by the present observation is
the comparison of the O/H ratio between the H\,{\sc i} and the
H\,{\sc ii} regions. We find $\log$(O\,{\sc i}/H\,{\sc
i})=$-4.7^{+0.8}_{-0.6}$ (2-$\sigma$), in agreement with the
O/H ratio measured in the H\,{\sc ii} region: $\log$(O/H)$_{\rm
HII}$=$-4.83\pm$0.03 (Izotov et al.~\cite{Izotov1999}). If the
H\,{\sc i} gas is more metal-deficient than the H\,{\sc ii}
regions, the difference is at most 0.5~dex (at 2~$\sigma$).

From the analysis of the same {\sl FUSE} data, Aloisi et al.
(\cite{Aloisi2003a}, \cite{Aloisi2003b}) found a significantly
different value of the O\,{\sc i}/H\,{\sc i} ratio :
$\log($O\,{\sc i}/H\,{\sc i})=$-5.4\pm$0.3. if true, this would
indicate that the H\,{\sc i} gas is more metal-deficient than the
H\,{\sc ii} regions. However their O\,{\sc i} column density
($\log (N_{\rm cm^{-2}}$(O\,{\sc i}))=15.98$\pm$0.26) is slightly
below our 2-sigma limit. While this discrepancy is to be
clarified, a likely source of a systematic error 
could be their use of the very
saturated O\,{\sc i} line at $\lambda_0=1039.23$~\AA , which is
contaminated by a terrestrial airglow.

Interestingly enough, O\,{\sc i}/H\,{\sc i} ratios similar to ours
are found in two other blue compact dwarf galaxies : Mrk~59
($\log$(O\,{\sc i}/H\,{\sc i})=$-5.0\pm$0.3 while
$\log$(O/H)$_{\rm HII}$=$-4.011\pm$0.003, Thuan et
al.~\cite{Thuan2002}), and SBS~0035-052 ($\log$(O\,{\sc i}/H\,{\sc
i})=$-5.0\pm$1.1 with $\log$(O/H)$_{\rm HII}$=$-4.70$). This may
suggest that the metallicity of the surrounding neutral envelope
of these blue galaxies is not related to the metallicity of the
H\,{\sc ii} regions, which could have been self-enriched in metals
by current or previous star formation bursts. However this may be
coincidental; the error bars are large and Mrk~59 is up to now the
only case for which a significant difference between the H\,{\sc
i} and H\,{\sc ii} regions has been found.

\subsection{N/O ratio}

The measured N\,{\sc i} column density gives $\log$(N\,{\sc
i}/O\,{\sc i})=$-2.4^{+0.6}_{-0.8}$ (2-$\sigma$), which is
not consistent with a N/O ratio of about $-1.5$ expected at low
metallicity for primary nitrogen. Even if some N\,{\sc i} is
ionized into N\,{\sc ii}, Jenkins et al. (\cite{Jenkins2000})
showed that the efficiency of photoionization arising from hot
stars is at most 0.15~dex. Furthermore, nitrogen is only slightly
depleted onto dust grains.

A possible explanation would be the ionization of N\,{\sc i} by
40-80~eV photons arising from the recombinaison of He\,{\sc ii},
as suggested by Jenkins et al. (\cite{Jenkins2000}) in the case of
the local interstellar medium. A similar nitrogen deficiency is
also observed in about 40\% of damped Lyman\,$\alpha$ systems
which have apparently not yet reached the full primary nitrogen
enrichment (Pettini et al.~\cite{Pettini2002}). A possible
explanation would be that the time delay for release of primary
nitrogen is longer when metal abundances are lower (see Pettini et
al.~\cite{Pettini2002} and references therein). Determination of
N\,{\sc i} and N\,{\sc ii} column densities in other blue compact
dwarf galaxies is certainly needed to solve this puzzling low
nitrogen content.

\subsection{Ar/O ratio}

Oxygen, silicon, and argon are $\alpha$-chain elements produced by
the same massive stars. As a consequence the Ar/O ratio is found
almost constant in H\,{\sc ii} regions of blue compact dwarf
galaxies (Thuan et al.~\cite{Thuan1995}) or low surface-brightness
dwarf galaxies (Van Zee et al.~\cite{vanZee1997}) and close to the
solar value: $\log$(Ar\,{\sc i}/O\,{\sc i})$\approx -2.3$.

In the neutral gas of \iz\ we measured $\log$(Ar\,{\sc i}/O\,{\sc
i})$= -3.15^{+0.6}_{-0.85}$, {\it a priori} lower than expected.
Although the \emph{statistical} uncertainty in the Ar/O ratio is
dominated by the much larger errors on the oxygen column density,
Levshakov et al. (\cite{Levshakov2001}) raised the possibility of
\emph{systematic} errors on argon due to the complexity of the
lines of sight. They showed that the column densities of argon,
silicon, and iron can be significantly underestimated with the
assumption of a single Voigt profile. Assuming {\it a priori} that
the argon abundance is the same in the H\,{\sc i} and the H\,{\sc
ii} regions, they found a larger argon column density due to the
saturation the argon line at $\lambda_0$=1048~\AA . However, these
authors fitted only the line at $\lambda_0$=1048~\AA; they did not
fit the fainter line at $\lambda_0$=1066~\AA. Even with the column
density given by Levshakov et al. (\cite{Levshakov2001}) ($\log
(N_{\rm cm^{-2}}$(Ar\,{\sc i}))$\sim 14.46$), and a simple line of
sight, this line at $\lambda_0$=1066~\AA\ is not saturated. It
appears difficult to have a complex line of sight which leads to
an underestimation by a factor of 10 in the equivalent width of
this line, and simultaneously an O\,{\sc i} line which goes to
almost zero at $\lambda_{\rm obs}$=978.9~\AA\ (see
Fig.~\ref{fit_OI} and Sect.~\ref{Multiple lines of sight}).

A more likely explanation of the apparently low Ar\,{\sc
i}/O\,{\sc i} ratio is a significant fraction of Ar\,{\sc i}
ionized into Ar\,{\sc ii}. This ionization is supposed to be
responsible for the low  Ar\,{\sc i}/H\,{\sc i} ratio measured in
the local interstellar medium, lower than the cosmic value of Ar/H
(Jenkins et al.~\cite{Jenkins2000}). In general, the argon is more
strongly ionized than hydrogen because the cross section for the
photoionization of Ar\,{\sc i} is very high. Jenkins et al.
(\cite{Jenkins2000}) showed that the deficiency in Ar\,{\sc i}
caused by the photoionization can be down to $\sim-$0.4~dex for a
shielding against the ionization photons of $N$(H\,{\sc i})$\sim
10^{18}$~cm$^{-2}$.

Nonetheless, the large error bars on the oxygen column density
could itself partially solve the issue of the apparent low
Ar\,{\sc i}/O\,{\sc i} ratio.

As a summary, the Ar\,{\sc i}/O\,{\sc i} ratio provides two
different clues. First, a hint towards a significant ionization of
argon as observed in the local interstellar medium. Second, a hint
of the favored oxygen column density within the measured
2-$\sigma$ interval. For example a value around $\log (N_{\rm
cm^{-2}}$(O\,{\sc i}))$\sim 16.3$ would give $\log$(Ar\,{\sc
i}/O\,{\sc i})$\sim -2.8$ and $\log$(Si\,{\sc ii}/O\,{\sc i})$\sim
-1.5$. The Ar\,{\sc i}/O\,{\sc i} ratio would be thus consistent
with the standard value, with a partial ionization. The Si/O ratio
would be the same as the value measured in the \iz\ H\,{\sc ii}
region (Izotov et al.~\cite{Izotov1999}).

\section{Conclusion}

We have measured column densities of H\,{\sc i}, N\,{\sc i},
O\,{\sc i}, Ar\,{\sc i}, Si\,{\sc ii}, Fe\,{\sc ii} in the neutral
gas of \iz . We obtain the following results:

\begin{enumerate}
\item The measured O\,{\sc i}/H\,{\sc i} ratio is consistent with
the O/H ratio as observed in the H\,{\sc ii} regions. We find
$\log$(O\,{\sc i}/H\,{\sc i})=$-4.7^{+0.85}_{-0.6}$ (2-$\sigma$)
which is also consistent with the O/H ratios $\sim
-5$ measured with {\sl FUSE} in the H\,{\sc i} regions of other
blue compact galaxies. If the oxygen is depleted in the H\,{\sc i}
region compared to the H\,{\sc ii} regions, the depletion is at
most 0.5~dex in \iz .
\item The measured N\,{\sc i}/O\,{\sc i} ratio is $\log$(N\,{\sc
i}/O\,{\sc i})=$-2.4^{+0.6}_{-0.8}$ (2-$\sigma$). This is
lower than the value of $-1.5$ expected for the N/O ratio with
primary nitrogen. The determination of the N\,{\sc ii} column
density is needed to discriminate between an ionization of N\,{\sc
i} or a possible nitrogen deficiency.
\item The neutral argon is also apparently underabundant. A
significant ionization into Ar\,{\sc ii} appears likely. The small
error bars in the determination of Ar\,{\sc i} and Si\,{\sc ii}
column densities favor the lower values of the large error bars in
the O\,{\sc i} column density, around $\log (N_{\rm
cm^{-2}}$(O\,{\sc i}))$\sim 16.3$.

\end{enumerate}

\begin{acknowledgements}
This work has been done using the profile fitting procedure 
developed by M.~Lemoine and the {\sl FUSE} French Team. The data
were obtained for the Guaranteed Time Team by the NASA-CNES-CSA
{\sl FUSE} mission operated by the Johns Hopkins University.
Financial support for French participants has been provided by
CNES. We warmly thank G.~\"Ostlin, J.~Lequeux and J.M.~Mas-Hesse
for fruitful discussions. We thank the anonymous referee for
helpful remarks.

\end{acknowledgements}

\end{document}